# On the Asymptotic Bias of the Diffusion-Based Distributed Pareto Optimization

Reza Arablouei [a], Kutluyıl Doğançay [b], Stefan Werner [c], and Yih-Fang Huang [d]

[a] Commonwealth Scientific and Industrial Research Organisation (CSIRO), Pullenvale, QLD, Australia
[b] School of Engineering, University of South Australia, Mawson Lakes, SA, Australia
[c] Department of Signal Processing and Acoustics, School of Electrical Engineering, Aalto University, Espoo, Finland
[d] Department of Electrical Engineering, University of Notre Dame, Notre Dame, IN, USA

**Abstract**: We revisit the asymptotic bias analysis of the distributed Pareto optimization algorithm developed based on the diffusion strategies. We propose an alternative way to analyze the asymptotic bias of this algorithm at small step-sizes and show that the asymptotic bias descends to zero with a linear dependence on the largest step-size parameter when this parameter is sufficiently small. In addition, through the proposed analytic approach, we provide an expression for the small-step-size asymptotic bias when a condition assumed jointly on the combination matrices and the step-sizes does not strictly hold. This is a likely scenario in practice, which has not been considered in the original paper that introduced the algorithm. Our methodology provides new insights into the inner workings of the diffusion Pareto optimization algorithm while being considerably less involved than the small-step-size asymptotic bias analysis presented in the original work. This is because we take advantage of the special eigenstructure of the composite combination matrix used in the algorithm without calling for any eigenspace decomposition or matrix inversion.

## 1. Introduction

Consider a connected network of $N$ nodes with processing and learning capabilities where the nodes can exchange information with their neighbors. Each node $k$ has its own individual convex cost function $J_k(w)$, which is minimized by the $M \times 1$ vector $w_k^o$. The minimizers $w_k^o$, $k = 1, ..., N$, are not necessarily the same. Therefore, finding a Pareto-optimal solution, denoted by the $M \times 1$ vector $w^o$, is of interest. The vector $w^o$ is taken as the minimizer of an aggregate cost function that is the equally-weighted sum of all individual cost functions, i.e.,

$$J^{\text{glob}}(w) = \sum_{k=1}^{N} J_k(w). \tag{1}$$

The Pareto-optimality of $w^o$ stems from the fact that any drift from $w^o$ to decrease any of the individual cost functions will result in an increase in at least one other individual cost function [1]-[3].

In [4], it is shown that the diffusion strategies for distributed optimization developed in [5] can be successfully employed to estimate the Pareto-optimal solution $w^o$ in a fully distributed manner, i.e., through only in-network processing and local interactions. In diffusion strategies, nodes collaborate by sharing their local estimates solely with their immediate neighbors. These strategies can be implemented on ad-hoc networks without warranting any fusion center, central processor, or hierarchy. This can particularly make them robust to possible topological changes in the network including link/node failures [4]-[18], [25]-[28].

The general algorithm that embodies both adapt-then-combine and combine-then-adapt variants of diffusion strategies for distributed optimization is described by the following recursive equations:

$$\phi_{k,i-1} = \sum_{l=1}^{N} a_{1,lk} w_{l,i-1} \tag{2}$$

$$\psi_{k,i} = \phi_{k,i-1} - \mu_k \sum_{l=1}^{N} c_{lk} \nabla_w J_l(\phi_{k,i-1}) \tag{3}$$

$$w_{k,i} = \sum_{l=1}^{N} a_{2,lk} \psi_{l,i}. \tag{4}$$

Here, $w_{k,i}$ is the local estimate of $w^o$ at node $k$ and iteration $i$, $\phi_{k,i-1}$ and $\psi_{k,i}$ are intermediate estimates, $\mu_k$ is the step-size parameter of node $k$, and $\nabla_w J_l(\cdot)$ is the gradient of $J_l(\cdot)$ with respect to $w$. Moreover, $a_{1,lk}$, $c_{lk}$, and $a_{2,lk}$ are the combination coefficients, which can be considered as $(l,k)$th entries of the combination matrices $A_1$, $C$, and $A_2$, respectively. They are required to satisfy

$$A_1^T \mathbb{1} = \mathbb{1}, A_2^T \mathbb{1} = \mathbb{1}, C\mathbb{1} = \mathbb{1},$$

$$a_{1,lk} = 0, a_{2,lk} = 0, c_{lk} = 0 \text{ if } l \notin \mathcal{N}_k$$

where $\mathbb{1}$ is the $N \times 1$ all-ones vector and $\mathcal{N}_k$ denotes the closed neighborhood of node $k$ that includes node $k$ as well [5]. Therefore, $A_1$ and $A_2$ are left-stochastic and $C$ is right-stochastic.

The algorithm (2)-(4) reduces to the adapt-then-combine diffusion strategy by choosing $A_1 = I_N$ and $A_2 = A$ and to the combine-then-adapt diffusion strategy by choosing $A_1 = A$ and $A_2 = I_N$ where $I_N$ is the $N \times N$ identity matrix. Apart from [4], this algorithm has effectively been used for distributed Pareto optimization in various scenarios and applications such as in the works of [19]-[23].

In the performance analysis presented in [4], it is shown that the asymptotic bias of the algorithm (2)-(4), when applied to distributed Pareto optimization, can be arbitrarily decreased by reducing the value of the step-size parameters with a rate of descent that is linear in the largest step-size parameter. This is verified in [4] by employing the Jordan canonical decomposition of the matrix $A_2^T A_1^T$ and the blockwise matrix inversion. In this paper, we prove the same relationship between the values of the asymptotic bias and the largest step-size parameter albeit via an approach that is considerably simpler than that of [4]. Moreover, we consider the case when a joint condition on the combination matrices and the step-sizes assumed in [4] (see Assumption 3 ahead) may be violated. This case has not been studied in [4]. We calculate the small-step-size asymptotic bias in this non-ideal situation to provide new insights into the performance of the diffusion-based Pareto optimization algorithm. In our analysis, we directly exploit the special eigenstructure of $A_2^T A_1^T$, which is assumed to be a right-stochastic and primitive matrix [4], and do not employ any matrix decomposition or inversion. We corroborate our theoretical findings through a simulated numerical example.

Note that we have adopted the same notation as in [4] to better elucidate our contributions. In addition, for the clarity of exposition, we repeat the equations and definitions given in [4] that are used in our arguments.

## 2. Asymptotic Bias

The following assumption on the cost functions is adopted in [4]:

*Assumption 1*: The Hessian matrix of the cost function of every node $k$, denoted by $\nabla_w^2 J_k(w)$, is bounded from below and from above, i.e., there exist $\lambda_{k,\min} \geq 0$ and $\lambda_{k,\max} > 0$ such that, for each $k = 1, \ldots, N$, we have

$$\lambda_{k,\min} I_M \leq \nabla_w^2 J_k(w) \leq \lambda_{k,\max} I_M \tag{5}$$

with

$$\sum_{l=1}^{N} c_{lk} \lambda_{l,\min} > 0. \tag{6}$$

Inequalities (5) mean that the eigenvalues of $\nabla_w^2 J_k(w)$ are upper and lower bounded by $\lambda_{k,\max}$ and $\lambda_{k,\min}$, respectively. Note that, in (5), the operator $\leq$ indicates that the difference of the matrices on its right- and left-hand sides is positive semi-definite. Assumption 1 guarantees that the aggregate cost $J^{\text{glob}}(w)$ defined by (1) is strongly convex. When $C = I_N$, (5) entails the strong convexity of the individual costs. Strong convexity is commonly assumed in the study of optimization techniques as it ensures that the Hessian matrix is not close-to-singular or ill-conditioned.

Under Assumption 1 and the following condition for all step-sizes

$$0 < \mu_k < \frac{2}{\sum_{l=1}^{N} c_{lk} \lambda_{l,\max}}. \tag{7}$$

Theorem 1 of [4] states that, after running a sufficiently large number of iterations of the algorithm (2)-(4) minimizing (1), the estimate of every node $k$ will converge to a unique fixed-point, denoted by $w_{k,\infty}$.

Provided that the algorithm converges, its asymptotic bias is calculated in Section III.D of [4] as

$$\widetilde{w}_\infty = [I_{MN} - \mathcal{A}_2^T(I_{MN} - \mathcal{M}\mathcal{R}_\infty)\mathcal{A}_1^T]^{-1} \mathcal{A}_2^T \mathcal{M} \mathcal{C}^T g^o \tag{8}$$

where

$$\widetilde{w}_{k,\infty} = w^o - w_{k,\infty} \text{ for } k = 1, \ldots, N,$$

$$\widetilde{w}_\infty = \begin{bmatrix} \widetilde{w}_{1,\infty} \\ \vdots \\ \widetilde{w}_{N,\infty} \end{bmatrix},$$

$$\mathcal{A}_1 = A_1 \otimes I_M, \mathcal{A}_2 = A_2 \otimes I_M, \mathcal{C} = C \otimes I_M,$$

$$\Omega = \text{diag}\{\mu_1, \cdots, \mu_N\}, \mathcal{M} = \Omega \otimes I_M,$$

$$\begin{aligned} H_{lk,\infty} &= \int_0^1 \nabla_w^2 J_l\left(w^o - t\sum_{l=1}^N a_{1,lk}\widetilde{w}_{l,\infty}\right) dt \\ &\approx \nabla_w^2 J_l(w^o), \end{aligned} \tag{9}$$

$$\begin{aligned} \mathcal{R}_\infty &= \sum_{l=1}^N \text{blkdiag}\{c_{l1}H_{l1,\infty}, \cdots, c_{lN}H_{lN,\infty}\} \\ &\approx \sum_{l=1}^N \text{blkdiag}\{c_{l1}\nabla_w^2 J_l(w^o), \cdots, c_{lN}\nabla_w^2 J_l(w^o)\}, \end{aligned}$$

$$g^o = \begin{bmatrix} \nabla_w J_1(w^o) \\ \vdots \\ \nabla_w J_N(w^o) \end{bmatrix}.$$

## 3. Small-Step-Size Asymptotic Bias

### 3.1. Previous Approach

Define

$$\mu_{\max} = \max\{\mu_1, \ldots, \mu_N\}, \quad \Omega_0 = \frac{\Omega}{\mu_{\max}}$$

and consider the following additional assumptions originally made in [4]:

*Assumption 2*: The composite combination matrix $A_1 A_2$ is left-stochastic and primitive. Thus, it has a unique eigenvalue at one with all other eigenvalues being smaller than one. Denote the right eigenvector of $A_1 A_2$ corresponding to its eigenvalue at one by $\theta$, which is normalized so that its entries add up to one, i.e., $\theta^T \mathbb{1} = 1$.

*Assumption 3*: It holds that

$$\theta^T A_2^T \Omega_0 C^T = c_0 \mathbb{1}^T \tag{10}$$

where $c_0$ is a constant.

One way to satisfy (10) is to make $A_1 A_2$ doubly-stochastic and the step-sizes of all nodes equal. Another way is to set $C = I_N$ and construct $A_1 A_2$ such that

$$\theta = (\sum_{k=1}^N \mu_k^{-1})^{-1} \begin{bmatrix} \mu_1^{-1} \\ \vdots \\ \mu_N^{-1} \end{bmatrix}.$$

Theorem 3 of [4] states that, under Assumptions 2 and 3, when the step-sizes are sufficiently small, i.e., as the largest step-size $\mu_{\max}$ approaches zero, the Euclidean norm of the asymptotic bias, $\widetilde{w}_\infty$, is of order $O(\mu_{\max})$. This theorem is proved in Appendix B of [4] by showing that

$$\lim_{\mu_{\max} \to 0} \frac{\|\widetilde{w}_\infty\|}{\mu_{\max}} = \xi$$

where $\xi$ is a constant independent of $\mu_{\max}$. The proof relies on the Jordan canonical decomposition of the matrix $A_2^T A_1^T$, i.e.,

$$A_2^T A_1^T = YJY^{-1}$$

while partitioning $J$, $Y$, and $Y^{-1}$ as

$$J = \begin{bmatrix} 1 & 0 \\ 0 & J_0 \end{bmatrix}, \quad Y^{-1} = \begin{bmatrix} \theta^T \\ Y_R \end{bmatrix}, \quad Y = [\mathbb{1}, Y_L],$$

and the blockwise matrix inversion formula, i.e.,

$$\begin{bmatrix} P & Q \\ U & V \end{bmatrix}^{-1} = \begin{bmatrix} P^{-1} + P^{-1}QSUP^{-1} & -P^{-1}QS \\ -SUP^{-1} & S \end{bmatrix}$$

where $S = (V - UP^{-1}Q)^{-1}$.

*3.2. New Approach*

In this subsection, we show that, as $\mu_{\max}$ approaches zero, the asymptotic bias, $\widetilde{w}_\infty$, tends to zero while being in the order of $O(\mu_{\max})$. To realize this, we merely exploit the special eigenstructure of $\mathcal{A}_2^T \mathcal{A}_1^T$ and do not employ any matrix decomposition or inversion.

To begin with, define

$$\mathcal{M}_0 = \Omega_0 \otimes I_M = \frac{\mathcal{M}}{\mu_{\max}},$$

$$\mathcal{X} = I_{MN} - \mathcal{A}_2^T \mathcal{A}_1^T,$$

$$\mathcal{Y} = \mathcal{A}_2^T \mathcal{M}_0 \mathcal{R}_\infty \mathcal{A}_1^T.$$

Then, rewrite (8) as

$$\widetilde{w}_\infty = \mu_{\max}(\mathcal{X} + \mu_{\max}\mathcal{Y})^{-1} \mathcal{A}_2^T \mathcal{M}_0 \mathcal{C}^T g^o. \tag{11}$$

From (5) and (9), we have

$$\lambda_{k,\min} I_M \leq H_{lk,\infty} \leq \lambda_{k,\max} I_M$$

and consequently

$$\mu_k \sum_{l=1}^N c_{lk} \lambda_{l,\min} I_M \leq \mu_k \sum_{l=1}^N c_{lk} H_{lk,\infty} \leq \mu_k \sum_{l=1}^N c_{lk} \lambda_{l,\max} I_M. \tag{12}$$

Under (6) and (7), (12) leads to

$$0 < \mu_k \sum_{l=1}^N c_{lk} H_{lk,\infty} < 2 I_M$$

or

$$-I_M < I_M - \mu_k \sum_{l=1}^N c_{lk} H_{lk,\infty} < I_M,$$

which implies that the spectral radius of $I_{MN} - \mathcal{M}\mathcal{R}_\infty$ is smaller than one. Thus, as $\mathcal{A}_1$ and $\mathcal{A}_2$ are left-stochastic, in virtue of Lemma D.6 of [6], the spectral radius of $\mathcal{A}_2^T(I_{MN} - \mathcal{M}\mathcal{R}_\infty)\mathcal{A}_1^T$ is also smaller than one. This means, $\mathcal{X} + \mu_{\max}\mathcal{Y}$ is positive-definite and invertible. Therefore, we can define the following limit matrix:

$$\mathcal{Z} = \lim_{\mu_{\max} \to 0} \mu_{\max}(\mathcal{X} + \mu_{\max}\mathcal{Y})^{-1}. \tag{13}$$

Calculating the limit in (13) is not trivial since $\mathcal{X}$ is rank-deficient. However, considering

$$(\mathcal{X} + \mu_{\max}\mathcal{Y})^{-1}\mathcal{X} = (\mathcal{X} + \mu_{\max}\mathcal{Y})^{-1}(\mathcal{X} + \mu_{\max}\mathcal{Y} - \mu_{\max}\mathcal{Y})$$
$$= I_{MN} - \mu_{\max}(\mathcal{X} + \mu_{\max}\mathcal{Y})^{-1}\mathcal{Y}$$

and

$$\mathcal{X}(\mathcal{X} + \mu_{\max}\mathcal{Y})^{-1} = (\mathcal{X} + \mu_{\max}\mathcal{Y} - \mu_{\max}\mathcal{Y})(\mathcal{X} + \mu_{\max}\mathcal{Y})^{-1}$$
$$= I_{MN} - \mu_{\max}\mathcal{Y}(\mathcal{X} + \mu_{\max}\mathcal{Y})^{-1}$$

as well as the definition of $\mathcal{Z}$ in (13), we can verify that

$$\mathcal{Z}\mathcal{X} = \lim_{\mu_{\max} \to 0} \mu_{\max}(\mathcal{X} + \mu_{\max}\mathcal{Y})^{-1}\mathcal{X}$$
$$= \lim_{\mu_{\max} \to 0} \mu_{\max}(I_{MN} - \mathcal{Z}\mathcal{Y}) = 0 \tag{14}$$

and

$$\begin{aligned}\mathcal{X}\mathcal{Z} &= \lim_{\mu_{\max}\to 0}\mu_{\max}\mathcal{X}(\mathcal{X}+\mu_{\max}\mathcal{Y})^{-1} \\ &= \lim_{\mu_{\max}\to 0}\mu_{\max}(I_{MN}-\mathcal{Y}\mathcal{Z})=0.\end{aligned} \quad (15)$$

Equations (14) and (15) entail that the columns of $\mathcal{Z}$ are in the kernel (nullspace) of $\mathcal{X}$ and the rows of $\mathcal{Z}$ are in the cokernel (left nullspace) of $\mathcal{X}$ [24]. Note that $\mathcal{X}$ can be written as

$$\begin{aligned}\mathcal{X} &= I_{MN}-A_2^T A_1^T\otimes I_M \\ &= (I_N - A_2^T A_1^T)\otimes I_M.\end{aligned}$$

Since $A_2^T A_1^T$ has a unique eigenvalue at one with the corresponding right and left eigenvectors $\mathbb{1}$ and $\theta^T$, respectively, $I_N - A_2^T A_1^T$ has a unique eigenvalue at zero, a kernel that is spanned by $\mathbb{1}$, and a cokernel that is spanned by $\theta^T$. Consequently, $\mathcal{X}$ has an eigenvalue at zero with multiplicity $M$ and the kernel and cokernel of $\mathcal{X}$ are spanned by the columns of $\mathbb{1}\otimes I_M$ and the rows of $\theta^T\otimes I_M$, respectively. Thus, due to (14), (15), and the abovementioned property of $\mathcal{X}$, $\mathcal{Z}$ can be factorized as

$$\mathcal{Z} = (\mathbb{1}\otimes I_M)\mathcal{D}(\theta^T\otimes I_M) \quad (16)$$

where $\mathcal{D}$ is some nonzero $M\times M$ matrix. Note that the $MN\times MN$ matrix $\mathcal{Z}$ has a rank of $M$.

Taking the limit at $\mu_{\max}\to 0$ on both sides of (11) followed by the substitution of (13) and (16) results in

$$\begin{aligned}\lim_{\mu_{\max}\to 0}\widetilde{w}_\infty &= (\mathbb{1}\otimes I_M)\mathcal{D}(\theta^T\otimes I_M)\mathcal{A}_2^T\mathcal{M}_0\mathcal{C}^T g^o \\ &= (\mathbb{1}\otimes I_M)\mathcal{D}(\theta^T A_2^T\Omega_0 C^T\otimes I_M)g^o.\end{aligned} \quad (17)$$

Since $w^o$ is the minimizer of the aggregate cost function $J^{\text{glob}}(w)$ defined by (1), we have

$$(\mathbb{1}^T\otimes I_M)g^o = 0. \quad (18)$$

Hence, in view of (18) and under Assumption 3, we get

$$\begin{aligned}(\theta^T A_2^T\Omega_0 C^T\otimes I_M)g^o &= c_0(\mathbb{1}^T\otimes I_M)g^o \\ &= 0.\end{aligned} \quad (19)$$

Substituting (19) into (17) gives

$$\lim_{\mu_{\max}\to 0}\widetilde{w}_\infty = 0.$$

It can also be concluded that since $\mathcal{Z}\mathcal{X}$ and $\mathcal{X}\mathcal{Z}$ descend to zero linearly in $\mu_{\max}$ as $\mu_{\max}$ approaches zero [see (14) and (15)], the descent of the asymptotic bias, $\widetilde{w}_\infty$, is also linear in $\mu_{\max}$, i.e., the asymptotic bias is of order $O(\mu_{\max})$, when $\mu_{\max}$ is sufficiently small.

*3.3. Without Assumption 3*

The analysis of [4] is limited to the case when Assumption 3 holds. This assumption is rather restrictive as it requires that either
- $A_1$ and $A_2$ are doubly-stochastic and the step-sizes of all nodes are equal or
- an elaborate joint design of the combination matrices and the step-sizes is performed.

In practice, we may choose to use non-ideal but convenient values for the combination matrices or step-sizes, which do not satisfy Assumption 3, at the cost of some possibly tolerable increase in the asymptotic bias. Hence, it is of practical importance to predict the small-step-size asymptotic bias of the diffusion Pareto optimization algorithm when Assumption 3 does not hold. One might rightfully argue that the asymptotic bias in this case can be calculated using (11) or its equivalent (8). However, the small-step-size asymptotic bias, which is the limit of the asymptotic bias as the step-sizes tend to zero, cannot be straightforwardly computed from (11) unless it is evaluated for several values of $\mu_{\max}$ and its asymptotic trend is observed. Therefore, (11) is not as useful and informative as a direct expression for calculating the small-step-size asymptotic bias in the absence of Assumption 3. In this subsection, we derive this expression.

From (13) and (16), we have

$$\lim_{\mu_{\max}\to 0}\mu_{\max}(\mathcal{X}+\mu_{\max}\mathcal{Y})^{-1} = (\mathbb{1}\otimes I_M)\mathcal{D}(\theta^T\otimes I_M). \quad (20)$$

Using

$$\mathcal{X}(\mathbb{1}\otimes I_M) = 0,$$

we can show that

$$\mu_{\max}^{-1}(\mathcal{X} + \mu_{\max}\mathcal{Y})(\mathbb{1}\otimes I_M) = \mu_{\max}^{-1}\mathcal{X}(\mathbb{1}\otimes I_M) + \mathcal{Y}(\mathbb{1}\otimes I_M)$$
$$= \mathcal{Y}(\mathbb{1}\otimes I_M). \tag{21}$$

Multiplying (20) by (21) from the right yields

$$\mathbb{1}\otimes I_M = (\mathbb{1}\otimes I_M)\mathcal{D}(\theta^T\otimes I_M)\mathcal{Y}(\mathbb{1}\otimes I_M)$$
$$= \mathbb{1}\otimes\mathcal{D}(\theta^T\otimes I_M)\mathcal{Y}(\mathbb{1}\otimes I_M),$$

which implies that

$$\mathcal{D}(\theta^T\otimes I_M)\mathcal{Y}(\mathbb{1}\otimes I_M) = I_M$$

or equivalently

$$\mathcal{D} = [(\theta^T\otimes I_M)\mathcal{Y}(\mathbb{1}\otimes I_M)]^{-1}. \tag{22}$$

Equation (22) can also be written as

$$\mathcal{D} = \left(\sum_{k=1}^{N} z_k \sum_{l=1}^{N} c_{lk}\nabla_w^2 J_l(w^o)\right)^{-1}$$

where $z_k$ is the $k$th entry of the vector

$$z = \Omega_0 A_2 \theta. \tag{23}$$

Note that all entries of $z$ are non-negative and, under Assumption 1, $(\theta^T\otimes I_M)\mathcal{Y}(\mathbb{1}\otimes I_M)$ is indeed invertible.

Substituting (22) and (23) into (17) gives the small-step-size asymptotic bias for the more general case that is not restricted to Assumption 3 as:

$$\lim_{\mu_{\max}\to 0}\widetilde{w}_\infty = (\mathbb{1}\otimes I_M)[(\theta^T\otimes I_M)\mathcal{Y}(\mathbb{1}\otimes I_M)]^{-1}(z^T C^T\otimes I_M)g^o$$
$$= (\mathbb{1}\otimes I_M)\left(\sum_{k=1}^{N} z_k \sum_{l=1}^{N} c_{lk}\nabla_w^2 J_l(w^o)\right)^{-1}\left(\sum_{k=1}^{N} z_k \sum_{l=1}^{N} c_{lk}\nabla_w J_l(w^o)\right). \tag{24}$$

Equation (24) specifies that the small-step-size asymptotic bias of all nodes converge to the same value of

$$\lim_{\mu_{\max}\to 0}\widetilde{w}_{k,\infty} = [(\theta^T\otimes I_M)\mathcal{Y}(\mathbb{1}\otimes I_M)]^{-1}(z^T C^T\otimes I_M)g^o.$$

This is evidently because of the left-stochasticity of $A_1 A_2$ and cannot be immediately deduced from (11). Note that, considering (18), when Assumption 3 holds, the right-hand side of (24) equals zero, which is consistent with the results of Section 3.2.

## 4. Simulations

In this section, we examine our theoretical results through a set of numerical experiments. We consider a network of $N = 50$ nodes with an arbitrary topology where each node is connected to four other nodes on average. We set $M = 4$ and choose the cost function of node $k$ as

$$J_k(w) = \|X_k w - y_k\|^2$$

where $X_k$ is a $6\times 4$ matrix and $y_k$ is a $6\times 1$ vector. We draw the entries of $X_k$ of $y_k$ for all nodes from a Gaussian distribution with unit variance and zero mean. The goal of each node is to find the minimizer of the aggregate cost function

$$J^{\text{glob}}(w) = \sum_{k=1}^{N}\|X_k w - y_k\|^2,$$

which is the global least-squares solution given by

$$w^o = \left(\sum_{k=1}^{N} X_k^T X_k\right)^{-1}\left(\sum_{k=1}^{N} X_k^T y_k\right).$$

We use both adapt-the-combine ($A_1 = I_N$, $A_2 = A$) and combine-then-adapt ($A_1 = A$, $A_2 = I_N$) variants of the diffusion Pareto optimization algorithm of (2)-(4) to estimate $w^o$ at the nodes. We consider both cases of all nodes having and not having a common step-size. In the latter case, we set the step-size of one of the nodes to $\mu_{\max}$ and draw the step-sizes of the other nodes from a uniform distribution on the interval $[\mu_{\max}/2, \mu_{\max}]$. Since the combination matrix $C$ is required to be right-stochastic, we construct its transpose $C^T$ using the averaging rule or the relative-degree rule. To construct $A$, we use the averaging, the relative-degree, or the Metropolis rule [7]. Using the averaging rule, the $(l,k)$th entry of $A$ is calculated as

$$a_{lk} = \begin{cases} \dfrac{1}{n_k} & l \neq k \\ 1 - \displaystyle\sum_{l \in \mathcal{N}_k \backslash \{k\}} a_{lk} & l = k \end{cases}$$

where $n_k = |\mathcal{N}_k|$ is the connectivity degree of node $k$, i.e., the number of its neighbors including itself. With the relative-degree or Metropolis rules, the $(l,k)$th entry of $A$ is calculated as

$$a_{lk} = \frac{n_l}{\sum_{m \in \mathcal{N}_k} n_m}$$

or

$$a_{lk} = \begin{cases} \dfrac{1}{\max\{n_l, n_k\}} & l \neq k \\ 1 - \displaystyle\sum_{l \in \mathcal{N}_k \backslash \{k\}} a_{lk} & l = k, \end{cases}$$

respectively. Using the Metropolis rule for creating $A$ makes $A_1 A_2$ doubly-stochastic and thus results in the satisfaction of Assumptions 3 when the step-sizes of all nodes are equal. However, the uniform and relative-degree rules do not lead to the satisfaction of Assumption 3 since the combination matrix created by either of these rules is not necessarily doubly-stochastic.

In Figs. 1-4, we plot the experimental values of the square-norm of the asymptotic bias against the largest step-size. We also include the square-norm of the theoretical values of small-step-size asymptotic bias predicted by (24). Figs. 1 and 3 correspond to the adapt-then-combine strategy while Figs. 2 and 4 correspond to the combine-then-adapt strategy. In addition, Figs. 1 and 2 are for the cases of having unequal step-sizes and Figs. 3 and 4 are for the cases of having equal step-sizes at all nodes. In Figs. 3(b) and 4(b), $A_1 A_2$ is doubly-stochastic and the step-sizes are equal hence Assumption 3 holds. Therefore, as anticipated by our analysis as well as [4], the square-norm of the asymptotic bias descends linearly in $\mu_{\max}^2$. We do not include the theoretical values of the small-step-size asymptotic bias in these figures as they are zero ($-\infty$ dB).

An excellent agreement between the theoretical predictions and the experimental observations is evident from Figs. 1-4. Moreover, we appreciate that, to achieve relatively low estimation errors, it is not necessary to fulfill Assumption 3. The combination matrices or the step-sizes that do not satisfy Assumption 3 still lead to decent estimation accuracies in the experiments of Figs. 1-4. Therefore, calculating the small-step-size asymptotic bias when Assumption 3 does not hold is of practical significance.

## 5. Conclusion

We studied the asymptotic bias of the diffusion Pareto optimization algorithm in the small-step-size regime from a different perspective by exploiting the special eigenstructure of the composite combination matrix without utilizing any matrix decomposition or inversion. Consequently, we provided a proof for asymptotic unbiasedness of the algorithm at small step-sizes that is appreciably less tortuous than the original proof given in [4]. Moreover, building on our new methodology, we derived a compact expression for the small-step-size asymptotic bias of the algorithm when the combination matrices or the step-sizes are not restricted by an assumption made in [4]. This expression provides new insights into the performance of the algorithm, specifically, when the combination matrices or step-sizes are not selected ideally.

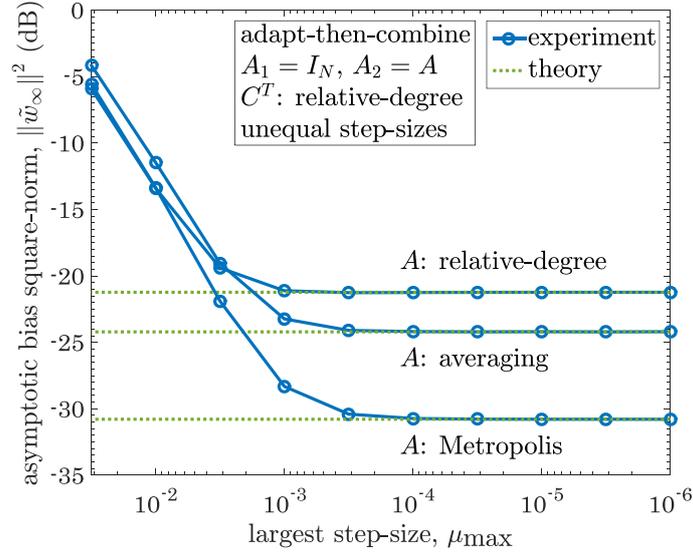

Fig. 1. The square-norm of the asymptotic bias versus the largest step-size together with the square-norm of the theoretically-predicted small-step-size asymptotic bias. The adapt-then-combine strategy ($A_1 = I_N$ and $A_2 = A$) is used with different choices of the combination matrix $A$. The relative-degree rule is used for the combination matrix $C^T$ and the step-sizes of the nodes are *not* equal.

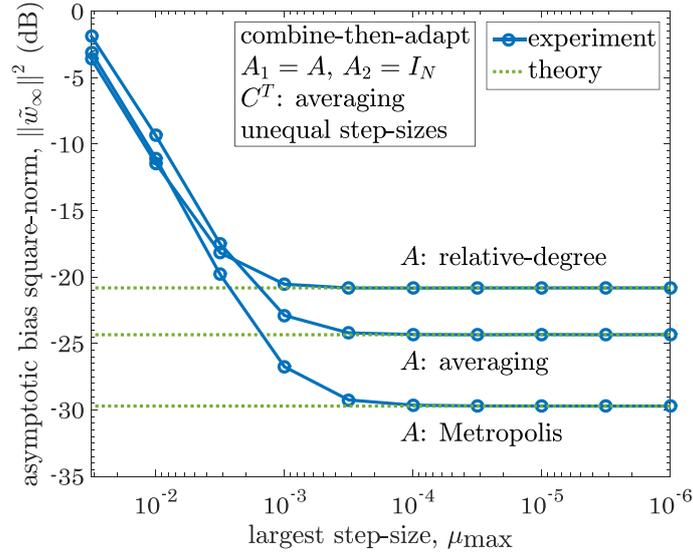

Fig. 2. The square-norm of the asymptotic bias versus the largest step-size together with the square-norm of the theoretically-predicted small-step-size asymptotic bias. The combine-then-adapt strategy ($A_1 = A$ and $A_2 = I_N$) is used with different choices of the combination matrix $A$. The averaging rule is used for the combination matrix $C^T$ and the step-sizes of the nodes are *not* equal.

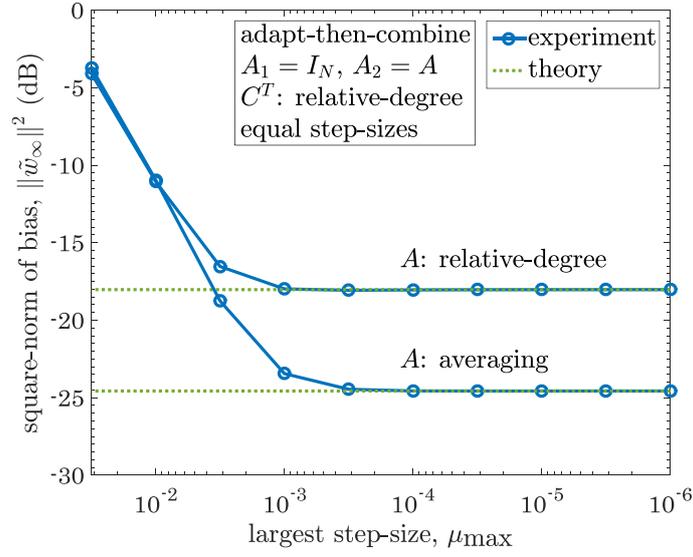

(a)

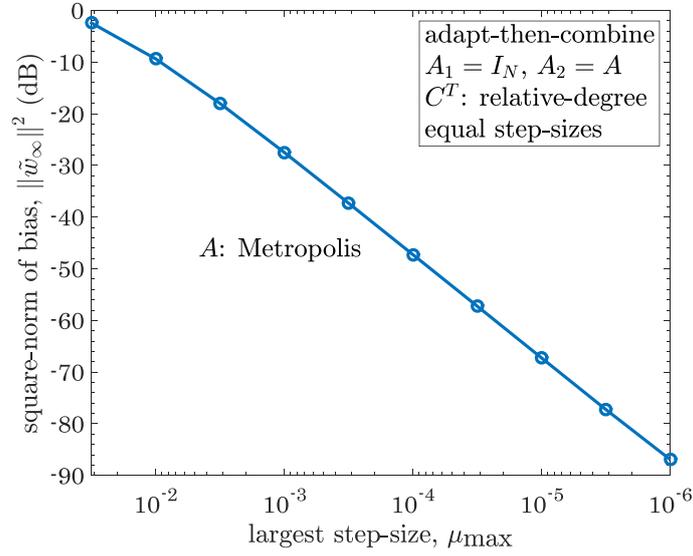

(b)

Fig. 3. The square-norm of the asymptotic bias versus the largest step-size together with the square-norm of the theoretically-predicted small-step-size asymptotic bias. The adapt-then-combine strategy ($A_1 = I_N$ and $A_2 = A$) is used with different choices of the combination matrix $A$. The relative-degree rule is used for the combination matrix $C^T$ and the step-sizes of the nodes are equal.

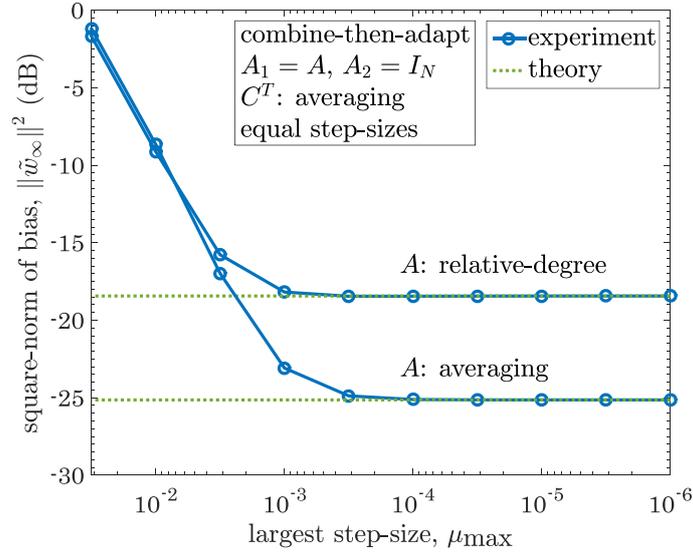

(a)

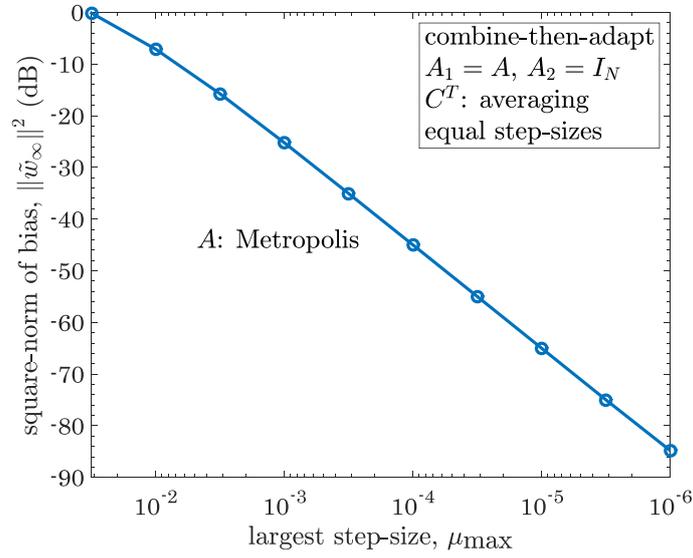

(b)

Fig. 4. The square-norm of the asymptotic bias versus the largest step-size together with the square-norm of the theoretically-predicted small-step-size asymptotic bias. The combine-then-adapt strategy ($A_1 = A$ and $A_2 = I_N$) is used with different choices of the combination matrix $A$. The averaging rule is used for the combination matrix $C^T$ and the step-sizes of the nodes are equal.